\documentstyle[twocolumn,prl,aps]{revtex}
\begin{document}
\twocolumn[\hsize\textwidth\columnwidth\hsize\csname
@twocolumnfalse\endcsname

\title{DISCRETE SCALE INVARIANCE IN TURBULENCE?}

\author{D. SORNETTE$^{1,2}$}
\address{$^1$ LPMC, CNRS UMR6622 and Universit\'e de Nice-Sophia Antipolis\\  B.P. 71, 
06108 NICE Cedex 2, France\\
$^2$ IGPP and Department of ESS,
UCLA\\ Los Angeles, CA 90095-1567}

\date{\today}
\maketitle
\begin{abstract} 
Based on theoretical argument and experimental evidence, we conjecture
that structure functions of turbulent times series exhibit
log-periodic modulations decorating their power law dependence.
In order to provide ironclad experimental evidence, we stress the need
for novel methods of averaging and propose a 
novel ``canonical'' averaging scheme for the analysis of structure factors 
of turbulent flows. The strategy is to determine the scale $r_c$ at which
the dissipation rate is the largest in a given turn-over time series. This
specific scale $r_c$ translates into a specific ``phase'' in the logarithm
of the scale which, when used as the origin, allows one to phase up the
different measurements of a structure factor $S_p(r) = A_p (\bar\epsilon
r)^{p/3}$ in different turn-over time realizations. We expect, as in
Laplacian growth and in rupture, that the log-periodic
oscillations will be reinforced by this canonical averaging. 
Demonstrating unambiguously the presence of log-periodicity and
thus of discrete scale invariance (DSI) in turbulent time-series would
provide an important step towards a direct demonstration of the Kolmogorov
cascade or at least of its hierarchical imprint. 
\end{abstract}
\vskip 1cm

to appear in the Proceedings of the Seventh European Turbulence Conference (ETC-7), 
June 30-July 3 (1998) (Published by Kluwer, U. Frisch, editor)
\vskip 1cm

% ] marks end of single wide column text region of
% \twocolumn[...text...]
]

\narrowtext

Scaling laws provide a quantification of
the complexity of turbulent flows.
For instance, the second order longitudinal structure function has the form 
$\langle (v_r)^2 \rangle = C_K (\bar\epsilon r)^{2/3}$, according to Kolmogorov 1941
\cite{Frisch} 
where $r$, which lies in the inertial range,
is the scale at which velocity differences are measured, and $\bar\epsilon$ is the mean
rate of energy dissipation per unit mass. 
Dimensional analysis shows that $\langle (v_r)^2 \rangle = 
(\bar\epsilon r)^{2/3} F(\Re, r/L)$, where
$F(x,y)$ is a universal function to be determined, $L$ is the
external or integral scale. Kolmogorov's assumption is that, for the Reynolds 
number $\Re \rightarrow
\infty$ and $r/L \rightarrow 0$, $F(x,y)$ goes to a constant $C_K$.
This is the so-called complete similarity of the first kind \cite{Barenblatt} with respect to 
the variables $\Re$ and $r/L$.

The existence of the limit of $F(\Re, r/L \to 0)$
has first been questioned by L.D. Landau and A.M. Obukhov, on the basis of the
existence of intermittency -- fluctuations of the energy dissipation rate about its mean value
$\bar\epsilon$. Indeed, Barenblatt's classification leads to the possibility
of an {\it incomplete similarity\/} in the variable $r/L$. This
would require the absence of a finite
limit for $F(\Re, r/L)$ as $r/L\rightarrow 0$, and leads in the simplest case to the form
$\langle (v_r)^2 \rangle = C_K (\bar\epsilon r)^{2/3} \left(\frac{r}{L}\right)^{\alpha}$,
where $\alpha$ is the so-called intermittency exponent, believed to be
small and positive. If $\alpha$ is real, this corresponds to a similarity of the second kind
\cite{Barenblatt}.
Incomplete self-similarity 
\cite{Goldenfeld,Dubrulle} may stem from a possible $\Re$-dependence of the exponents.
The case where $\alpha$ is complex, leading to
$\langle (v_r)^2 \rangle = C_K (\bar\epsilon r)^{2/3} 
\left(\frac{r}{L}\right)^{\alpha_R}~\cos[\alpha_I \log(r/L)]$,
could be termed a similarity of the third kind, characterized by the absence of limit for 
$F(\Re, r/L)$ and accelerated (log-periodic) oscillations.
To our knowledge,  Novikov has been the first to point out in 1966 that structure functions
in turbulence should contain log-periodic oscillations \cite{Novikov}. His argument was that
if an unstable eddy in a turbulent flow typically breaks up into two
or three smaller eddies, but not into $10$ or $20$ eddies, then one can suspect
the existence of a prefered scale factor, hence the log-periodic
oscillations.
They have been repeatedly observed
but do not seem to be stable and depend on the nature of the
global geometry of the flow and recirculation \cite{Frisch,Anselmet}
as well as the analyzing procedure.

The theory of complex exponents and log-periodicity has advanced significantly \cite{revue} in 
the last few years. Complex exponents reflect a discrete scale invariance (DSI), i.e.
the fact that dilational symmetry occurs only under magnification under 
special factors, which are arbitrary powers $\lambda^n$ of a prefered scaling ratio
$\lambda$. Complex exponents have been studied in the eighties in relation
to various problems of physics embedded in hierarchical systems. In the context
of turbulence, shell models construct explicitely a discrete scale 
invariant set of equations whose solutions are marred by unwanted log-periodicities.
Only recently has it been realized that discrete scale invariance and its
associated complex exponents may appear ``spontaneously'' in euclidean
systems, i.e. without the need for a pre-existing hierarchy. Systems that
have been found to exhibit self-organized DSI are Laplacian growth models \cite{DLA},
rupture in heterogeneous systems \cite{Anifrani}, earthquakes \cite{SorSam},
  animals \cite{SS} (a generalization of
percolation) among many other systems. In addition,
general field theoretical arguments \cite{SS}
indicate that complex exponents are to be expected generically for 
out-of-equilibrium and/or quenched disordered systems. This together with 
Novikov's argument suggest to revisit log-periodicity in
turbulent signals. Demonstrating
unambiguously the presence of log-periodicity and thus of DSI in turbulent
time-series would provide an important step towards a direct demonstration of the
Kolmogorov cascade or at least of its hierarchical imprint.
For partial indications of log-periodicity in turbulent data, 
we refer the reader to fig. 5.1 ~p.58 and fig. 8.6~ p.128 of Ref.
\cite{Frisch}, fig.3.16~ p. 76 of Ref.\cite{Arneodo}, fig.1b of Ref.\cite{Tcheou} and 
fig. 2b of Ref. \cite{Castaing}.

It is a common observation that
the oscillations, if any, ``move'' when changing the length
of the signal over which the averaging is carried out. They thus have 
the aspect of noise. However, previous numerical
simulations on Laplacian growth models \cite{DLA,finitezeiz} and 
renormalization group 
calculations \cite{SS} have taught us that 
the presence of noise modifies the phase in the
log-periodic oscillations in a sample specific way leading to a ``destructive
interference'' upon averaging. In the turbulence context, 
we propose that one realization corresponds approximately to 
a signal measured over one turn-over time scale $L/v_L$. In contrast to this
sample specific phase dependence, we stress that the prefered scaling ratio $\lambda$ 
has universal properties.

It is thus
important to carry out an analysis on each sample realization separately,
without averaging, as has been demonstrated to work for other systems 
\cite{DLA}. An enticing alternative is to introduce
a new averaging scheme that does not destroy the oscillations.
The standard
averaging procedure, that we could term ``Grand canonical'' \cite{Pazmandi},
is known to introduce spurious sample-to-sample fluctuations
of relative amplitude proportional to $L^{-d/2}$ in $d$ dimensions. In contrast, 
the concept of ``canonical'' averaging \cite{Pazmandi}
consists in identifying, for each realization, the corresponding specific value 
of the critical control parameter $K_c^R$. The natural control parameter
then becomes $\Delta = (K-K_c^R)/K_c^R$ and the averaging can then be
performed over the different samples with the same $\Delta$. 
This should then lead in principle to a
``rephasing'' of the log-oscillations. This ``canonical averaging'' has been
demonstrated for log-periodic signatures of the acoustic emission precursors prior to rupture
and in Laplacian growth models \cite{finitezeiz}.

We propose to adapt this ``canonical'' averaging scheme to the analysis of structure functions 
of turbulent flows. There are probably several possible schemes to implement it. Let us
suggest here one based on the energy dissipation rate. The strategy is to look for 
a reference quantity that is specific to a given turn-over time realization. 
For critical phenomena, a
natural candidate is the susceptibility whose maximum determines the sample specific
critical point location $K_c^R$ \cite{Pazmandi,finitezeiz}. For 
turbulence, we suggest to determine the scale $r_c$ at which the dissipation rate is the largest in
a given turn-over time series. 
This can be derived by a direct measurement of 
the velocity gradient at small scales or from the third-order structure function
$S_3(r) = - {4 \over 5} \bar\epsilon r$,
which obeys the exact four-fifth Kolmogorov law under a set of assumptions \cite{Frisch}.
This specific scale $r_c$ translates into a specific ``phase'' in the logarithm of the scale
which, when used as the origin, allows one to phase up the different measurements of a
structure function $S_p(r) = A_p (\bar\epsilon r)^{p/3}$ in different turn-over time realizations.
We expect, as in Laplacian growth models and in rupture, that the log-periodic oscillations
will be reinforced by this canonical averaging.

What could be the mechanism that creates these
characteristic scales? There are undoubtly the integral length $L$ and the scales associated
to the dissipation range. But what could produce an approximate geometrical series of scales in
the inertial range? We conjecture two possible routes. The first one is inspired from 
a recent discovery that the continuous nonlinear Einstein 
partial differential equations of general relativity
in the presence of a scalar field self-interacting through gravitation may generate 
a log-periodic spectrum of black hole masses with develop according to a log-periodic
self-similar time dynamics \cite{Blackhole}. The mechanism 
might result from the existence of a limit cycle in the renormalization group description of
a field close to the negative density limit (in turbulence, could this be obtained from
a negative effective viscosity?). The other route is that scale invariant equations
that present an instability at finite wavevector $k$ decreasing with the field
amplitude may generate naturally a spectrum of internal scales. An example is 
${\partial v \over \partial t} = -2v{\partial^2 v \over \partial x^2} 
- v^2 {\partial^4 v \over \partial x^4}$.
This equation is scale invariant in the sense that if $v(t,x)$ is a solution, then 
$\gamma^2 v(t, \gamma x)$ is also a solution for arbitrary $\gamma$. A linear 
stability analysis shows that a mode $v_0 e^{\sigma t} e^{ikx}$ grows with 
$\sigma = 2v_0 k^2 - v_0^2 k^4$, i.e. the most unstable mode occurs at finite $k_{m.u.} = 
{1 \over \sqrt{v_0}}$. Thus, a finite characteristic scale appears that is
completely controlled by the amplitude of the field. Starting from an
approximate homogeneous level $v_0$, the instability produces a large
scale ${2\pi \over k_{m.u.}} = 2 \pi \sqrt{v_0}$. As dips in the field develop, the amplitude
there decreases and the corresponding instabilities will create smaller length scales, and so on.
Preliminary simulations \cite{Gil}
confirm this intuitive picture\,: the resulting DSI field is seen to 
result from a cascade of instabilities with characteristic wavelengths controlled by the 
amplitude.

I hope that these conjectural ideas will stimulate further works on
these fascinating log-periodic structures in turbulent signals.
I am grateful to U. Frisch, L. Gil, N. Goldenfeld, A. Johansen, 
A. Noullez and  G. Simms for discussions.


\begin{thebibliography}{99}

\bibitem{Frisch} Frisch, U. (1995) {\it Turbulence, the legacy of A.N. Kolmogorov},
Cambridge University Press.

\bibitem{Barenblatt}  Barenblatt, G. I. (1996)
 {\it Scaling, self-similarity, and intermediate asymptotics},
Cambridge University Press.
   
\bibitem{Goldenfeld} Barenblatt, G.I. \& Goldenfeld, N. (1995) {\it Phys. Fluids} {\bf 7},
pp.~3078-3082.
     
\bibitem{Dubrulle} Dubrulle, B., (1996) {\it J. Phys. France II} {\bf 6}, pp.~1825-1840.

\bibitem{Novikov} Novikov, E.A. (1966) {\it Dokl.Akad.Nauk SSSR} {\bf 168/6}, 
pp.~1279; (1990) {\it Phys.Fluids} A {\bf 2}, pp.~814--820.

\bibitem{Anselmet} Anselmet, F., Gagne, Y., Hopfinger, E.J. \& Antonia, R.A. 
(1984) {\it J. Fluid Mech.} {\bf 140}, pp.~63 

\bibitem{revue} Sornette, D. (1998) Discrete scale invariance and complex dimensions, 
{\it Physics
Reports}, in press (april 1998) (http://xxx.lanl.gov/abs/cond-mat/9707012).

\bibitem{DLA} Sornette, D., Johansen, A., Arn\'eodo, A., Muzy, J.-F. \& Saleur, H.
(1996) {\it Phys. Rev. Lett.} {\bf 76}, pp.~251--254; 
Huang, Y., Ouillon, G., Saleur, H. \& Sornette, D. (1997) 
{\it Phys. Rev.} E {\bf 55}, pp.~6433--6447.

\bibitem{Anifrani} Anifrani, J.-C., Le Floc'h, C., Sornette, D. \&
Souillard, B. (1995) {\it J.Phys.I France} {\bf 5}, pp.~ 631--638.

\bibitem{SorSam}  Sornette, D. \& Sammis, C.G. (1995) 
{\it J.Phys.I France} {\bf 5}, pp.~607--619;
Johansen, A.,  Sornette, D., Wakita, H., Tsunogai, U., Newman, W.I. \&
 Saleur, H. (1996) {\it J.Phys.I France} {\bf 6}, pp.~1391--1402.

\bibitem{SS} Saleur, H. \& Sornette, D. (1996) 
 {\it J.Phys.I France} {\bf 6}, pp.~327--355.

\bibitem{Arneodo} Arn\'eodo, A., Argoul, F., Bacry, E., Elezgaray, J. \& Muzy, J.-F. (1995)
{\it Ondelettess, multifractales et turbulences}, Diderot Editeur, Arts et Sciences

\bibitem{Tcheou} Tch\'eou, J.-M. \& Brachet, M.E. (1996) {\it J.Phys.II France} {\bf 6}, 
pp.~ 937--943.

\bibitem{Castaing} Castaing, B. (1997) 
in {\it Scale invariance and beyond}, eds. Dubrulle, B., Graner,
F. \& Sornette, D., EDP Sciences and Springer, pp.~ 225--234.

\bibitem{finitezeiz} A. Johansen \& D. Sornette (1998)
Evidence for discrete scale invariance by canonical averaging, preprint.

\bibitem{Pazmandi} Pazmandi, F., Scalettar, R.T. \& Zimanyi, G.T. (1997) 
{\it Phys. Rev. Lett.} {\bf 79}, pp.~5130--5133 (1997).

\bibitem{Blackhole} Choptuik, M.W. (1993) 
{\it Phys. Rev. Lett.} {\bf 70}, pp.~9--12.

\bibitem{Gil} Gil, L. \& Sornette, D. (1998), in preparation.

\end{thebibliography}
\end{document}